\documentstyle[twoside,11pt]{article}
\def\greaterthansquiggle{\raise.3ex\hbox{$>$\kern-.75em\lower1ex\hbox{$\sim$}}}
\def\lessthansquiggle{\raise.3ex\hbox{$<$\kern-.75em\lower1ex\hbox{$\sim$}}}
\newcommand{\beq}{\begin{equation}}
\newcommand{\eeq}{\end{equation}}
\newcommand{\beqa}{\begin{eqnarray}}
\newcommand{\eeqa}{\end{eqnarray}}
\newcommand{\beqan}{\begin{eqnarray*}}
\newcommand{\eeqan}{\end{eqnarray*}}
\newcommand{\ba}{\begin{array}}
\newcommand{\ea}{\end{array}}
\newcommand{\no}{\nonumber}

\newcommand{\ra}{\rightarrow}

\newcommand{\ve}{\varepsilon}

\newcommand{\D}{{\cal D}}

\newcommand{\cL}{{\cal L}}

\newcommand{\dfrac}{\displaystyle \frac}

\newcommand{\dsum}{\displaystyle \sum}

\def\nz{\ifmmode {I\hskip -3pt N} \else {\hbox {$I\hskip -3pt N$}}\fi}
\def\zz{\ifmmode {Z\hskip -4.8pt Z} \else
       {\hbox {$Z\hskip -4.8pt Z$}}\fi}
\def\qz{\ifmmode {Q\hskip -5.0pt\vrule height6.0pt depth 0pt
       \hskip 6pt} \else {\hbox
       {$Q\hskip -5.0pt\vrule height6.0pt depth 0pt\hskip 6pt$}}\fi}
\def\rz{\ifmmode {I\hskip -3pt R} \else {\hbox {$I\hskip -3pt R$}}\fi}
\def\cz{\ifmmode {C\hskip -4.8pt\vrule height5.8pt\hskip 6.3pt} \else
       {\hbox {$C\hskip -4.8pt\vrule height5.8pt\hskip 6.3pt$}}\fi}

\def\au{{\setbox0=\hbox{\lower1.36775ex%
\hbox{''}\kern-.05em}\dp0=.36775ex\hskip0pt\box0}}
\def\ao{{}\kern-.10em\hbox{``}}

\normalsize
\begin{document}
\bibliographystyle{plain}
\begin{titlepage}
\begin{flushright}
UWThPh-1993-23\\
DPUR 64\\
\today
\end{flushright}
\vspace{2cm}
\begin{center}
{\Large \bf Transition Amplitudes within the Stochastic Quantization
Scheme}\\[40pt]
H. H\"uffel  \\
Institut f\"ur Theoretische Physik \\
Universit\"at Wien \\
A-1090 Wien, Austria \\
email: A8241DAG@AWIUNI11.BITNET \\[5pt]
and \\[5pt]
H. Nakazato \\
Department of Physics \\
University of the Ryukyus \\
Okinawa 903-01, Japan \\
\centerline{\vtop{\hbox{email:B985039@SCI.U-RYUKYU.AC.JP}
      \hbox{\hphantom{email:}\& D52787@JPNKUDPC.BITNET}}}
\vfill
{\bf Abstract} \\
\end{center}

Quantum mechanical transition amplitudes are calculated within the
stochastic quantization scheme for the free nonrelativistic particle,
the harmonic oscillator and the nonrelativistic particle in a
constant magnetic field; we close with free Grassmann quantum
mechanics.
\vfill
\end{titlepage}

\renewcommand{\theequation}{\arabic{section}.\arabic{equation}}
\setcounter{equation}{0}
\section{Introduction}
Stochastic quantization [1] was introduced more than a decade ago as a
novel method for quantizing field theories and provides an intriguing
connection between statistical mechanics and quantum field theory
[2--3].

It is the aim of this paper to explore the somewhat more elementary
issue of how to calculate within the stochastic scheme the transition
amplitudes of ordinary quantum mechanics. Amusingly this problem has
passed practically unnoticed over the years (see however [4--5]) so we
would like to present it in a rather elementary way: We study the
nonrelativistic free particle, the harmonic
oscillator and the nonrelativistic particle in a constant magnetic field,
presenting a phase space as well as a configuration space formulation;
we close with the free Grassmann quantum mechanics [6].

At the beginning we recall basic facts on stochastic quantization:
Given some arbitrary functional $f$ of coordinates $x(t)$, its vacuum
expectation value can be obtained as the equilibrium limit $s \ra \infty$ of
the stochastic correlation function $\langle f[x;s]\rangle$
\beq
\langle f[x]\rangle_0 = \lim_{s \ra \infty} \langle f[x;s]\rangle
\eeq
which can be defined as
\beq
\langle f[x;s] \rangle = \int \D x f[x] P [x;s].
\eeq
Here $P[x;s]$ is a normalized (generalized) probability distribution,
obeying the Fokker--Planck equation
\beq
\frac{\partial}{\partial s} P[x;s] = \int_{-\infty}^\infty dt
\frac{\delta}{\delta x(t)} \left( \frac{\delta}{\delta x(t)} -
i \frac{\delta S}{\delta x(t)}\right) P[x;s].
\eeq
We work in real--time quantum mechanics and implicitly assume a convenient
$i \ve$--insertion in the action $S$ to guarantee the convergence of the
stochastic process\footnote{
  Actually we would like to avoid a general discussion of complex stochastic
  processes but we note that this prescription is sufficient for the case of
  the quadratic systems under consideration.}
 [7--10].
  Alternatively the stochastic correlations (1.2)
can be obtained by assigning to $x(t)$ an additional stochastic time
dependence $x(t) \ra x(t,s)$ via the Langevin equation
\beq
\frac{\partial x(t,s)}{\partial s} = i \frac{\delta S}{\delta x(t,s)} +
\eta(t,s).
\eeq
Its solution has to be inserted into the given functional $f$ and the average
over the white noise $\eta$, characterized by
\beq
\langle \eta(t,s) \eta(t',s')\rangle = 2 \delta(t-t') \delta(s-s')
\eeq
finally has to be worked out.

In quantum mechanics in contrast to the calculation of vacuum expectation
values one frequently is interested in the normalized transition
elements of time ordered products of operators between Heisenberg state
vectors $\vert x,t\rangle$ which are eigenstates of Heisenberg operators
$x(t)$ belonging to the eigenvalues $x$.
We therefore introduce new stochastic correlation
functions, taking care of the boundary values $x_i$ at $t_i$ and $x_f$
at $t_f$. We consider for example
\beq
\frac{\langle x_f,t_f\vert T(x(t_1)x(t_2))\vert x_i,t_i\rangle}
{\langle x_f,t_f \vert x_i,t_i\rangle} = \lim_{s \ra \infty}
\langle x(t_1,s) x(t_2,s)\rangle_{bc}
\eeq
and define in analogy to (1.2)
\beq
\langle x(t_1,s) x(t_2,s)\rangle_{bc} = \int_{x(t_i)=x_i, x(t_f) = x_f}
\D x \; x(t_1) x(t_2) P[x;s].
\eeq
$P$ now satisfies a Fokker--Planck equation (1.3) with the time integration
restricted to the interval $t \in [t_i,t_f]$; the normalization
condition reads $\langle 1 \rangle_{bc} = 1$.  It follows that the stationary
solution is modulo
given by $e^{iS}$ so that the standard path integral
representation emerges at the equilibrium limit of (1.7).

Throughout this
paper we prefer to use a Langevin equation approach
which seems favourable given the rich technical experience already
acquired in the case of field theories [2--3]; we will give the details of how
to implement the boundary conditions in the Langevin scheme in the next
section.  Furthermore our ultimate
goal (not yet addressed in this paper) will be the discussion of
constrained quantum mechanical systems and we expect intriguing advantages
of the Langevin approach over the rather complicated Fokker--Planck
formulation [4].

The central object on which
we would like to concentrate, however, is the transition amplitude
$\langle x_f,t_f \vert x_i,t_i\rangle$ itself, which is one of the fundamental
quantities in quantum theory.  Due to the normalization
condition for the probability density $P$ we cannot directly express the
transition amplitude in terms of it.
Stochastic expectation values calculated either with Fokker--Planck or
Langevin equation techniques are always
normalized automatically and it seemed difficult to find a way to reproduce
such quantities as transition amplitudes within the framework of the
stochastic quantization.
It is, however, indeed possible to relate
$\langle x_f,t_f\vert x_i,t_i\rangle$ to the normalized stochastic
expectation value of the Hamiltonian $H$.  This follows, as in the case of a
regular quantum mechanical system the Heisenberg state vectors fulfill
\beq
\vert x_i,t_i\rangle = T \exp[i \int_{t_0}^{t_i} H(t) dt] \vert x_i,t_0\rangle
\eeq
and
\beq
\frac{\partial}{\partial t_i} \vert x_i,t_i\rangle = i H(t_i) \vert x_i,t_i
\rangle.
\eeq
Therefore we get
\beq
\frac{\partial}{\partial t_i} \ln \langle x_f,t_f\vert x_i,t_i\rangle =
i \frac{\langle x_f,t_f\vert H(t_i)\vert x_i,t_i\rangle}
{\langle x_f,t_f \vert x_i,t_i\rangle}.
\eeq
The right hand side is a normalized expectation value of the Hamiltonian
at $t_i$, which implies that this quantity is obtainable as a limit of
the stochastic average $\langle H(t_i,s)\rangle_{bc}$.  Here note that for
conservative systems the stochastic average of the Hamiltonian is
$t$-independent at equilibrium $s=\infty$; for calculational simplicity we
evaluate it at the initial time $t=t_i$.  That is,
we have as a first step
\beq
\langle x_f,t_f \vert x_i,t_i\rangle = \tilde c \exp[ i \int^{t_i}
\lim_{s \ra \infty}
\langle H(t_i,s)\rangle_{bc} dt_i]
\eeq
with $\tilde c$ a constant independent of $t_i$.
For a transparent presentation we restrict ourselves to Hamiltonians
quadratic in the canonical variables. In this case we can separate off
very simply contributions from solutions to
the classical equations of motion and arrive at
\beqa
\langle x_f,t_f\vert x_i,t_i\rangle
&=& \tilde c \exp\left[i \int^{t_i} H_{\rm cl}(t_i) dt_i \right]
    \exp\left[i \int^{t_i} \lim_{s \ra \infty} \langle H_Q(t_i,s)
    \rangle_{bc} dt_i \right] \no \\
&=& c \exp[iS_{\rm cl}]
    \exp\left[i \int^{t_i} \lim_{s \ra \infty} \langle H_Q(t_i,s)
    \rangle_{bc} dt_i \right].
\eeqa
Here the classical action $S_{\rm cl}$ has appeared as a result of the famous
relation $\partial S_{\rm cl}/\partial t_i=H_{\rm cl}(t_i)$ [11]
and is composed of the classical path $x_{\rm cl}(t)$ as usual.
We remark that all the quantum contributions are contained in the second
exponential factor.

The constant $c$ in principle could depend on $x_f$
and $x_i$ which, however, is not
the case.  This follows from a similar
variational principle as above for $\langle x_f,t_f\vert x_i,t_i\rangle$
w.r.t.\ $x_f$ and $x_i$ and the relations $\partial S_{\rm cl}/\partial x_f=
p_{\rm cl}(t_f)$ and $\partial S_{\rm cl}/\partial x_i=-p_{\rm cl}(t_i)$.
So the constant $c$ is indeed $t_f$, $t_i$, $x_f$, $x_i$ independent and
can be fixed by
requiring the transition amplitude to approach a Dirac $\delta$--function
$\delta (x_f - x_i)$ in the limit of $t_i = t_f$ (compare also with
Ref.~[11]).

\renewcommand{\theequation}{\arabic{section}.\arabic{equation}}
\setcounter{equation}{0}
\section{Phase Space Formulation}
\subsection{General Procedure}
In the last section we related the transition amplitude to an
expectation value of the Hamiltonian,
so it appears appropriate to
rely on a phase space formulation of stochastic quantization [12].
We put
\beq
x(t) = x_{\rm cl}(t) + x_Q(t), \qquad
p(t) = p_{\rm cl}(t) + p_Q(t)
\eeq
and implement the boundary conditions for the quantum fields
$x_Q(t_i) = x_Q(t_f) = 0$  by the Fourier decompositions
\beqa
x_Q(t) &=& \sum_{n=1}^\infty x_n \sin \frac{n\pi}{T} (t-t_i), \\
p_Q(t) &=& \sum_{n=1}^\infty p_n \cos \frac{n\pi}{T} (t-t_i) + \frac{p_0}{2},
\qquad T = t_f - t_i.
\eeqa
Notice that no boundary conditions are imposed on the momentum variable
$p_Q(t)$ and that any function defined in $[t_i,t_i+T]$ can continuously
be extended to $[t_i-T,t_i+T]$ as an even function like the above $p_Q$.
Stochastic quantization proceeds by introducing the $s$-dependence for the
Fourier modes $x_n\ra x_n(s),p_n\ra p_n(s)$ according to the phase-space
Langevin equations
\beq
\frac{d}{ds} x_n = i\frac{\delta S_Q}{\delta x_n} + \xi_n, \qquad
\frac{d}{ds} p_n = i \frac{\delta S_Q}{\delta p_n} + \eta_n
\eeq
where the noises fulfill
\beqa
\langle \xi_n(s) \xi_m(s')\rangle &=& \langle \eta_n(s) \eta_m(s')\rangle
= 2 \delta_{nm} \delta(s-s'), \no \\
\langle \eta_n(s) \xi_m(s')\rangle &=& 0.
\eeqa
For the quadratic case (2.4) is explicitly solved by
\beq
\left(\ba{c} x_n\\ p_n\ea\right)(s)
=\int_0^s e^{iA_n(s-\sigma)}\left(\ba{c} \xi_n\\ \eta_n\ea\right)(\sigma)
 d\sigma
\eeq
where $A_n$ is a model-dependent matrix.  Working in Stratonovich calculus we
easily derive
\beqa
\langle x_n(s) \xi_m(s)\rangle &=& \langle p_n(s) \eta_m(s)\rangle =
\delta_{nm}, \no \\
\langle x_n(s) \eta_m(s)\rangle &=& \langle p_n(s) \xi_m(s)\rangle = 0
\eeqa
and find that correlations containing stochastic time derivatives
vanish in the equilibrium limit $s \ra \infty$
\beq
\left\langle x_n(s)\dfrac{dx_m(s)}{ds}\right\rangle
=\left\langle x_n(s)\dfrac{dp_m(s)}{ds}\right\rangle
=\left\langle p_n(s)\dfrac{dx_m(s)}{ds}\right\rangle
=\left\langle p_n(s)\dfrac{dp_m(s)}{ds}\right\rangle
\ra0.
\eeq
In the following examples, we avoid solving the Langevin equations
explicitly and instead extract the two-point correlations indirectly
by using (2.8).

\subsection{Free Particle}
The easiest example to discuss is the nonrelativistic free particle with the
Hamiltonian
\beq
H = \frac{p^2}{2M}.
\eeq
We have now
\beq
S_{\rm cl} = \frac{M}{2T} (x_f - x_i)^2, \qquad
S_Q = \sum_{n=1}^\infty \left( \frac{n\pi}{T} x_n p_n - \frac{p_n^2}{2M}
\right) \frac{T}{2} - p_0^2 \; \frac{T}{8M}
\eeq
and get
\beqa
\frac{dx_n}{ds} &=& \frac{in\pi}{2} p_n + \xi_n, \no \\
\frac{dp_n}{ds} &=& i \left(\frac{n\pi}{T} x_n - \frac{p_n}{M} \right)
\frac{T}{2} + \eta_n, \no \\
\frac{dp_0}{ds} &=& -i \frac{T}{4M} p_0 + \eta_0.
\eeqa
{}From $\left\langle p_m(s) \dfrac{dx_n(s)}{ds} \right\rangle$,
$\left\langle p_m(s) \dfrac{dp_0(s)}{ds} \right\rangle$ and
$\left\langle p_0(s) \dfrac{dp_0(s)}{ds} \right\rangle$ we immediately
find in the equilibrium limit
\beq
\langle p_m p_n\rangle = \lim_{s \ra \infty} \langle p_m(s) p_n(s)\rangle
= 0, \qquad \langle p_m p_0 \rangle = 0, \qquad
\langle p_0^2 \rangle = - \frac{4iM}{T},
\eeq
respectively. Furthermore,
\beq
\lim_{s \ra \infty} \langle H_Q(t_i,s)\rangle_{bc} = \lim_{s \ra \infty}
\frac{\langle p_Q^2(t_i,s)\rangle_{bc}}{2M} =
\frac{\langle p_0^2\rangle}{8M} = - \frac{i}{2T},
\eeq
so that
\beq
i \int^{t_i} \lim_{s \ra \infty} \langle H_Q(t_i,s)\rangle_{bc} dt_i
=-i \int^{T}(- \frac{i}{2T}) dT
=- \frac{1}{2}\ln{T}
\eeq
and finally
\beq
\langle x_f,t_f\vert x_i,t_i\rangle = c \frac{1}{\sqrt{T}} \; e^{iS_{\rm cl}},
\qquad
c = \sqrt{\frac{M}{2\pi i}}.
\eeq

\subsection{Harmonic Oscillator}
Another standard example is given by the harmonic oscillator
\beq
H = \frac{p^2}{2M} + \frac{1}{2} M \omega^2 x^2
\eeq
in which case
\beqa
S_{\rm cl} &=& \frac{\omega M}{2 \sin \omega T} [(x_i^2 + x_f^2)
\cos \omega T - 2x_i x_f], \\
S_Q &=& \sum_{n=1}^\infty \left(\frac{n\pi}{T} x_n p_n - \frac{p_n^2}{2M}
- \frac{1}{2} M \omega^2 x_n^2 \right)\frac{T}{2} - p_0^2 \frac{T}{8M}
\eeqa
so that
\beqa
\frac{dx_n}{ds} &=& i \left(\frac{n\pi}{T} p_n - M \omega^2 x_n\right)
\frac{T}{2} + \xi_n, \no \\
\frac{dp_n}{ds} &=& i \left(\frac{n\pi}{T} x_n - \frac{p_n}{M} \right)
\frac{T}{2} + \eta_n, \no \\
\frac{dp_0}{ds} &=& -i \frac{T}{4M} p_0 + \eta_0.
\eeqa
{}From $\left\langle p_n \dfrac{dp_0}{ds}\right\rangle$,
$\left\langle p_0 \dfrac{dp_0}{ds} \right\rangle$ we find in the
equilibrium limit
\beq
\langle p_n p_0\rangle = 0, \qquad
\langle p_0^2\rangle = - i \frac{4M}{T}
\eeq
and from $\left\langle p_m \dfrac{dx_n}{ds}\right\rangle$ combined with
$\left\langle p_m \dfrac{dp_n}{ds} \right\rangle$
\beq
\langle p_n p_m\rangle = \frac{2iM}{T} \frac{\delta_{nm}}
{\left( \dfrac{n\pi}{T\omega}\right)^2 - 1}.
\eeq
Therefore (recalling the boundary condition $x_Q(t_i)=0$)
\beq
\lim_{s \ra \infty} \langle H_Q(t_i,s)\rangle_{bc} = \frac{1}{2M}
\left(\sum_{n,m=1}^\infty \langle p_n p_m\rangle + \frac{\langle p_0^2
\rangle}{4} \right) = - \frac{i \omega}{2} \cot T\omega
\eeq
and the transition amplitude is obtained as
\beq
\langle x_f,t_f\vert x_i,t_i\rangle = c \frac{1}{\sqrt{\sin T\omega}}\,
e^{i S_{\rm cl}}, \qquad c = \sqrt{\frac{M\omega}{2\pi i}}.
\eeq

\subsection{Constant Magnetic Field}
A less trivial example of our approach is the nonrelativistic charged particle
in a constant magnetic field:
\beq
H = \frac{1}{2}(\vec p - \vec A)^2,  \qquad
\vec p = \left( \ba{c} p \\ q \\ r \ea \right), \qquad
\vec A = \left( \ba{c} 0 \\ Bx \\ 0 \ea \right), \qquad
\vec x = \left( \ba{c} x \\ y \\ z \ea \right).
\eeq
We split again
\beq
\vec x(t) = \vec x_{\rm cl}(t) + \vec x_Q(t), \qquad
\vec p(t) = \vec p_{\rm cl}(t) + \vec p_Q(t)
\eeq
and obtain
\beq
S_{\rm cl} = \frac{1}{2} \left\{ \frac{(z_f - z_i)^2}{T} + \frac{B}{2}
\cot \frac{BT}{2} [(x_f - x_i)^2 + (y_f - y_i)^2]
+ B(x_i + x_f)(y_f - y_i)\right\}
\eeq
as well as
\beqa
S_Q &=& - \frac{T}{8} \vec p_0{}^2 +
        \sum_{n=1}^\infty \left\{ \frac{n\pi}{T} \vec x_n \cdot\vec p_n -
        \frac{1}{2} \vec p_n{}^2 - \frac{B^2}{2}\vec x_n{}^2 \right.\no \\
&& \left. \mbox{} - \frac{BT}{\pi} \left[ x_n q_0 \frac{(-1)^n -1}{2n} -
\sum_{m=1}^\infty x_n q_m n \frac{(-1)^{n+m} -1}{n^2 - m^2} \right]
\right\}.
\eeqa
The Langevin equations read as
\beqa
\frac{d \vec x_n}{ds} &=& i\left(\vec p_n \frac{n\pi}{2} + \vec a_n\right)
+ \vec \xi_n \no \\
\frac{d \vec p_n}{ds} &=& i\left(\vec x_n \frac{n\pi}{2} - \vec p_n
\frac{T}{2} + \vec b_n\right) + \vec \eta_n \no \\
\frac{\vec p_0}{ds} &=& i \left( - \frac{T}{4} \vec p_0 + \vec c\right)
+ \vec \eta_0
\eeqa
where we have introduced
\beq
\vec a_n = \dfrac{BT}{2} \left( \ba{c}
                               \alpha_n \\ 0 \\ 0 \ea \right), \qquad
\vec b_n = \left( \ba{c}
                 0 \\ \beta_n \\ 0 \ea \right), \qquad
\vec c = \left( \ba{c}
               0 \\ \gamma \\ 0 \ea \right)
\eeq
with
\beqa
\alpha_n
&=&-Bx_n - q_0 \dfrac{(-1)^n -1}{n\pi}
   + \dfrac{2}{\pi} \dsum_{m=1}^\infty q_m n
     \dfrac{(-1)^{n+m} -1}{n^2 - m^2} \no \\
\beta_n
&=&-\dfrac{BT}{\pi} \dsum_{m=1}^\infty n
   \dfrac{(-1)^{n+m} -1}{n^2 - m^2} \no \\
\gamma
&=&-\dfrac{BT}{2\pi} \dsum_{m=1}^\infty x_m \dfrac{(-1)^m -1}{m}. \no
\eeqa
It follows from
$ \left\langle q_m \dfrac{dy_n}{ds}\right\rangle$,
$ \left\langle q_0 \dfrac{dy_n}{ds}\right\rangle$,
$ \left\langle p_n \dfrac{dp_0}{ds}\right\rangle$,
$ \left\langle r_m \dfrac{dz_n}{ds}\right\rangle$ and
$ \left\langle r_n \dfrac{dr_0}{ds}\right\rangle$ that
\beq
\langle q_n q_m\rangle = \langle q_n q_0 \rangle = \langle p_n p_0\rangle
= \langle r_n r_m\rangle = \langle r_n r_0\rangle = 0.
\eeq
Therefore
\beq
\lim_{s \ra \infty} \langle H_Q(t_i,s)\rangle_{bc} = \frac{1}{8}
\langle \vec p_0{}^2\rangle + \frac{1}{2} \sum_{n,m=1}^\infty
\langle p_n p_m\rangle .
\eeq
Using
$ \left\langle p_0 \dfrac{dp_0}{ds}\right\rangle$,
$ \left\langle r_0 \dfrac{dr_0}{ds}\right\rangle$ we have as previously
\beq
\langle p_0^2 \rangle = \langle r_0^2\rangle = - \frac{4i}{T}.
\eeq
{}From
$ \left\langle q_0 \dfrac{dp_n}{ds}\right\rangle$ we relate
$\langle q_0 p_n\rangle$ to $\langle x_n q_0\rangle$; then
$ \left\langle q_0 \dfrac{dx_n}{ds}\right\rangle$ connects
$\langle x_n q_0\rangle$ to $\langle q_0^2\rangle$ and from
$ \left\langle q_0 \dfrac{dq_0}{ds}\right\rangle$ we find
\beq
\langle q_0^2 \rangle = - i 2 B \cot \frac{BT}{2}.
\eeq
{}From
$ \left\langle p_m \dfrac{dp_n}{ds}\right\rangle$ we relate
$\langle x_n p_m\rangle$ to $\langle p_n p_m\rangle$;
from $ \left\langle p_m \dfrac{dx_n}{ds}\right\rangle$ relating
$\langle x_n p_m \rangle$, $\langle q_0 p_n\rangle$ and
$\langle q_0^2\rangle$ we obtain
\beq
\langle p_n p_{n'}\rangle = 2i \frac{B^2T}{\pi^2} \left[
\frac{\delta_{nn'}}{n^2 - \left( \frac{BT}{\pi} \right)^2}
- \frac{BT}{\pi^2}
\frac{(-1)^n -1}{n^2 - \left( \frac{BT}{\pi}\right)^2}
\frac{(-1)^{n'} -1}{n'{}^2 - \left( \frac{BT}{\pi}\right)^2}
\cot \frac{BT}{2}\right].
\eeq
Straightforwardly we finally arrive at
\beq
\lim_{s \ra \infty} \langle H_Q(t_i,s)\rangle_{bc} = -i \frac{B}{2}
\cot \frac{BT}{2} - \frac{i}{2T}
\eeq
and
\beq
\langle x_f,t_f\vert x_i,t_i\rangle = c \frac{1}{\sqrt{T \sin \frac{BT}{2}}}
e^{i S_{\rm cl}}, \qquad
c = \frac{B}{2} \left( \frac{B}{2\pi i}\right)^{3/2}.
\eeq

\renewcommand{\theequation}{\arabic{section}.\arabic{equation}}
\setcounter{equation}{0}
\section{Configuration Space Formulation}
\subsection{General Procedure}
Although as outlined in the last section the phase space formulation is
the natural one, it is for technical reasons preferable to rely
on the simpler configuration space formulation. If
\beq
H = \frac{p^2}{2M} + V(x), \qquad
\cL = \frac{M}{2} \dot x^2 - V(x)
\eeq
we have, of course,
\beq
\langle H_Q(t_i,s)\rangle_{bc} =
\left\langle p_Q(t_i,s) \frac{\partial x_Q}{\partial t} (t_i,s)
\right\rangle_{bc} -
\langle \cL_Q(t_i,s)\rangle_{bc}.
\eeq
Now
\beq
\langle \cL_Q(t_i,s)\rangle_{bc} = \frac{M}{2}
\left\langle \frac{\partial x_Q^2}{\partial t} (t_i,s)\right\rangle_{bc}
= \frac{M}{2} \lim_{t_1,t_2 \ra t_i} \partial_{t_1} \partial_{t_2}
\langle x_Q(t_1,s) x_Q(t_2,s)\rangle_{bc}
\eeq
where we have used Feynman's time splitting procedure to properly define the
kinetic energy contributions [11]. The transition to a pure
configuration space formulation is achieved by observing that generally
\beq
\frac{\partial p_Q}{\partial s}(t,s) = i
\left( \frac{\partial x_Q(t,s)}{\partial t} -
\frac{\partial H_Q}{\partial p_Q(t,s)} \right) + \xi(t,s).
\eeq
{}From
$\left\langle \dfrac{\partial x_Q}{\partial t}(t',s)
\dfrac{\partial p_Q}{\partial s} (t,s)\right\rangle_{bc}$
we find
\beq
\lim_{s \ra \infty} \left[
\left\langle \frac{\partial x_Q}{\partial t}(t',s)
p_Q(t,s)\right\rangle_{bc} - M
\left\langle \frac{\partial x_Q}{\partial t}(t',s)
\frac{\partial x_Q}{\partial t} (t,s)\right\rangle_{bc}\right] = 0
\eeq
and finally obtain
\beq
\lim_{s \ra \infty} \langle H_Q(t_i,s)\rangle_{bc} =\frac{M}{2}
\lim_{t_1,t_2 \ra t_i} \partial_{t_1} \partial_{t_2}
\lim_{s\ra\infty}\langle x_Q(t_1,s) x_Q(t_2,s)\rangle_{bc},
\eeq
which coincides with the very definition of the
Hamiltonian in the path-integral formulation [11] and can be
calculated entirely in configuration space.
We start from the regular Lagrangian $\cL$ and separate the coordinate
$x(t)$ into a classical and quantum part, the latter being Fourier
expanded as in (2.2). The configuration-space Langevin equations for
the Fourier modes are associated as in the original Parisi-Wu approach
and the transition amplitudes can be calculated using (1.12) and (3.6).
We remark that the transition to configuration space in the case of
the particle in a constant magnetic field arrives at the same
expression (3.6), even though it corresponds to a case of velocity-dependent
quantum mechanical potential.

\subsection{Free Particle}
To demonstrate the configuration space approach we start from the free
particle action
\beq
S = \int_{t_i}^{t_f}\! dt\, \frac{M}{2}\dot x^2, \qquad
S_Q = \sum_{n=1}^\infty \frac{n^2 \pi^2 M}{4T} x_n^2,
\eeq
and get
\beq
\frac{dx_n}{ds} = i \frac{n^2 \pi^2 M}{2T} x_n + \eta_n
\eeq
so that
\beq
\langle x_n x_m\rangle = \frac{2iT}{n^2 \pi^2 M} \delta_{nm}.
\eeq
{}From this we immediately find
\beq
\lim_{s \ra \infty} \langle x_Q(t_1,s) x_Q(t_2,s)\rangle_{bc} = \frac{i}{MT}
[t_f - \max (t_1,t_2)] [\min (t_1,t_2) - t_i]
\eeq
and have from (3.6)
\beq
\lim_{s \ra \infty} \langle H_Q(t_i,s)\rangle_{bc} = - \frac{i}{2T}.
\eeq

Here we have to keep in mind that we should first sum over Fourier modes
$n$ to obtain $\lim_{s\ra\infty}\langle x_Q(t_1,s)x_Q(t_2,s)\rangle_{bc}=
\langle x_Q(t_1)x_Q(t_2)\rangle_{bc}$, differentiate w.r.t.~$t_1$ and $t_2$
and finally take the limit $t_1,t_2 \ra t_i$.
This ordering corresponds to Feynman's time splitting procedure
[11], which avoids infinite contributions arising from the velocity
correlation at the same point.
If the order were reversed or changed, for example, if we differentiated
$x_Q(t)$ before the summation over $n$, we would have an infinite
contribution $\sum_{n=1}^\infty=\infty$, as is easily seen from (3.9).

\subsection{Harmonic Oscillator}
With
\beq
S = \int_{t_i}^{t_f} dt\left( \frac{M}{2} \dot x^2 - \frac{1}{2} M
\omega^2 x^2\right), \qquad
S_Q = \sum_{n=1}^\infty \frac{M\pi^2}{4T} \left[ n^2 -
\left( \frac{T \omega}{\pi}\right)^2 \right] x_n^2,
\eeq
we obtain now
\beq
\frac{dx_n}{ds} = i \frac{M \pi^2}{2T} \left[ n^2 -
\left( \frac{T\omega}{\pi}\right)^2 \right] x_n + \eta_n
\eeq
and
\beq
\langle x_n x_m\rangle = i \frac{2T}{M \pi^2}
\frac{\delta_{nm}}{n^2 - \left( \frac{T\omega}{\pi}\right)^2}.
\eeq
Furthermore we obtain
\beq
\lim_{s \ra \infty} \langle x_Q(t_1,s) x_Q(t_2,s)\rangle_{bc} =
\frac{i}{M \omega \sin T\omega}
\sin \omega(t_f - \max (t_1,t_2))\sin \omega(\min (t_1,t_2) - t_i)
\eeq
and get upon differentiation
\beq
\lim_{s \ra \infty} \langle H_Q(t_i,s)\rangle_{bc} = - \frac{i\omega}{2}
\cot T\omega .
\eeq

\subsection{Constant Magnetic Field}
Despite the simple looking form
\beq
\cL = \frac{\dot{\vec x}^2}{2} + \dot{\vec x} \cdot \vec A,
\eeq
$S_Q$ takes an involved expression
\beq
S_Q = \frac{\pi^2}{4T} \sum_{n,n'=1}^\infty (x_n,y_n,z_n)
\left( \ba{ccc}
D & M & 0 \\
-M& D & 0 \\
0 & 0 & D \ea \right)_{nn'}
\left( \ba{c} x_{n'} \\ y_{n'} \\ z_{n'} \ea \right)
\eeq
where
\beqa
D_{nn'} &=& n^2 \delta_{nn'}, \no \\
M_{nn'} &=& - \frac{(-1)^{n+n'} -1}{n^2 - n'{}^2} nn' \frac{2BT}{\pi^2}.
\eeqa
As the $z$-components decouple, we first investigate the $(x_n,y_n)$
contributions only. We start from the Langevin equations
\beq
\frac{\partial}{\partial s}
\left( \ba{c} x_n \\ y_n \ea \right) = \frac{i\pi^2}{2T}
\left( \ba{cc} D & M \\ - M & D \ea \right)_{nn'}
\left( \ba{c} x_{n'} \\ y_{n'} \ea \right) +
\left( \ba{c} \eta_n \\ \xi_n \ea \right)
\eeq
and multiply with $(x_{n''},y_{n''})$ to extract in the equilibrium limit
\beq
-\frac{i\pi^2}{2T} \left( \ba{cc} D & M \\ - M & D \ea \right)_{nn'}
\left\langle \left( \ba{c} x_{n'} \\ y_{n'} \ea \right)
(x_{n''},y_{n''}) \right\rangle =
\left( \ba{cc} {\bf 1}_{nn''} & 0 \\ 0 & {\bf 1}_{nn''} \ea \right).
\eeq
We eliminate the off-diagonal expectation values and find for $x_n$
(we remark that an identical expression holds for $y_n$)
\beq
-\frac{i\pi^2}{2T} (D + MD^{-1}M)_{nn'} \langle x_{n'} x_{n''}\rangle =
\delta_{nn''}
\eeq
which explicitly evaluates to
\beq
-\frac{i\pi^2}{2T} \left\{ \left[ n^2 - \left(\frac{BT}{\pi}\right)^2
\right] \delta_{nn'} + 2 \left(\frac{BT}{\pi^2}\right)^2
\frac{[(-1)^n -1][(-1)^{n'}-1]}{nn'}  \right\}
\langle x_{n'} x_{n''}\rangle = \delta_{nn''}.
\eeq
We find
\beq
\langle x_n x_{n'}\rangle = \frac{2iT}{\pi^2} \left\{
\frac{\delta_{nn'}}
{n^2 - \left(\frac{BT}{\pi}\right)^2}
- \frac{(BT)^3}{\pi^4}
\frac{[(-1)^n-1][(-1)^{n'}-1]}
{nn' \left[ n^2 - \left(\frac{BT}{\pi}\right)^2\right]
\left[ n'{}^2 - \left(\frac{BT}{\pi}\right)^2\right]}
\cot \frac{BT}{2}\right\}.
\eeq

  Therefore after the summations over $n$ and $n'$, we get the correlation
function
\beqa
\langle x_Q(t_1)x_Q(t_2)\rangle_{bc}
&=&\frac{i}{B\sin BT}\{\sin B(t_f-\max(t_1,t_2))\sin B(\min(t_1,t_2)-t_i)
                                                              \no \\
& &  -[\cos\frac{B}{2}(t_f+t_i-2t_1)-\cos\frac{BT}{2}]
     [\cos\frac{B}{2}(t_f+t_i-2t_2)-\cos\frac{BT}{2}]\},
                                                              \no \\
& &
\eeqa
from which follows
\beq
\lim_{t_1,t_2\ra t_i}\partial_{t_1}\partial_{t_2}\lim_{s\ra\infty}
\langle x_Q(t_1,s)x_Q(t_2,s)\rangle_{bc}=-\frac{iB}{2}\cot\frac{BT}{2}.
\eeq
Adding an identical $y_n$-contribution and the trivial $z_n$-contribution
we get in total as before
\beq
\lim_{s \ra \infty} \langle H(t_i,s)\rangle_{bc} = - \frac{iB}{2} \cot
\frac{BT}{2} - \frac{i}{2T}.
\eeq

\subsection{Nonrelativistic Grassmann Quantum Mechanics}
In this section we want to extend our method to deal with Grassmann
quantum systems. We study the simplest possible model, which is a free
nonrelativistic Grassmann particle described by [6]
\beq
\cL = \frac{M}{2} \left( \frac{d\Theta}{dt}\right)^2.
\eeq
Here we have introduced a Grassmann vector
\beq
\Theta = \{ \Theta^\alpha\}, \qquad \alpha = 1,2
\eeq
and defined an inner product
\beq
\Theta^2 = \Theta \cdot \Theta = \Theta^\alpha \Gamma_{\alpha\beta}
\Theta^\beta
\eeq
with the help of a metric tensor
\beq
\Gamma_{\alpha\beta} = - \Gamma^{\alpha\beta} = \left( \ba{cc}
0 & 1 \\ -1 & 0 \ea \right).
\eeq
Notice that (3.28) differs from Grassmann quantum mechanical
models which are inspired from particle physics and have kinetic terms
linear in the time derivative [13--15].
These models generally lead
to second class constraints, however. Having decided to restrict
ourselves in this paper to regular systems we  do not
include these models in our discussion.

We choose to work in the configuration space formulation and separate
the Grassmann variable into a classical and quantum part as before
\beq
\Theta(t) = \Theta_{\rm cl}(t) + \Theta_Q(t), \qquad
\Theta_Q(t) = \sum_{n=1}^\infty \Theta_n \sin \frac{n\pi}{T}(t - t_i)
\eeq
so that
\beq
S_{\rm cl} = \frac{M}{2} \frac{(\Theta_f - \Theta_i)^2}{T}, \qquad
S_Q = \sum_{n=1}^\infty \frac{n^2 \pi^2 M}{4T} \Theta_n^2.
\eeq
The definition of the Langevin equation for the Grassmann mode
$\Theta_n$ requires (similarly to the fermionic field theory case
[16]) the introduction of a kernel, which we choose to be
$\Gamma^{\alpha\beta}$
\beqa
\frac{d \Theta_n^\alpha}{ds}
&=& i \Gamma^{\alpha\beta}
\frac{\delta S_Q}{\delta \Theta_n^\beta} + \eta_n^\alpha
\no \\
&=& \frac{i n^2 \pi^2 M}{2T}
\Theta_n^\alpha + \eta_n^\alpha.
\eeqa
Here the noise fulfills
\beq
\langle \eta_n^\alpha(s) \eta_m^\beta(s')\rangle = 2 \delta_{nm}
\Gamma^{\alpha\beta} \delta(s-s')
\eeq
which guarantees the Grassmann anticommutativity.
{}From $\left\langle \Theta_m(s) \cdot \dfrac{d\Theta_n}{ds} (s)\right\rangle$
we find
\beq
\lim_{s \ra \infty} \langle \Theta_n(s) \cdot \Theta_m(s)\rangle =
- \frac{4iT}{n^2 \pi^2 M} \delta_{nm}
\eeq
where a crucial minus sign has appeared as a consequence of the definition
of the inner product with $\Gamma^{\alpha\beta}$. We easily deduce
\beq
\lim_{s \ra \infty} \langle \Theta_Q(t_1,s) \cdot\Theta_Q(t_2,s)\rangle_{bc} =
- \frac{2i}{MT} [t_f - \max (t_1,t_2)][\min (t_1,t_2)-t_i]
\eeq
so that after differentiation
\beq
\lim_{s \ra \infty} \langle H_Q(t_i,s)\rangle_{bc} = \frac{i}{T}
\eeq
and the desired transition amplitude becomes
\beq
\langle \Theta_f t_f\vert \Theta_i t_i\rangle = c T e^{i S_{\rm cl}}, \qquad
c =  \frac{i}{M}.
\eeq
Note that the constant $c$ was fixed in this example by requiring
the transition amplitude to approach a Berezin delta function in the equal
time limit $t_i = t_f$ [6].

\section{Conclusions}
In this paper we have filled a somewhat historical gap in the applications
of stochastic quantization and calculated for the nonrelativistic free
particle, the harmonic oscillator, the nonrelativistic particle in a constant
magnetic field and the free Grassmann particle the
quantum mechanical transition amplitudes.  Our main observation is the
possibility to relate the nonnormalized transition amplitudes with the
normalized exectation values within the stochastic quantization scheme taking
care of the boundary conditions.

  Our procedure can equally well be applied to Euclidean quantum mechanics,
in which case the corresponding partition functions become straightforwardly
calculable.

  It remains a challenge to apply our scheme to constrained systems.

\section*{Acknowledgements}
  One (H.N.) of the authors would like to express his gratitude to Saverio
Pascazio and Paolo Cea for helpful discussions.  He also wishes to thank all
members at the Institut f\"ur Theoretische Physik, Universt\"at Wien for warm
hospitality, where this joint work started.
Both authors appreciate the inspiring atmosphere of the
Madeira 1993 conference ``Stochastic Analysis
and Applications in Physics", during which this work was completed.

\end{document}